\documentclass[prb,twocolumn,showpacs,preprintnumbers,amsmath,amssymb]{revtex4}

\usepackage{graphicx}
\usepackage{dcolumn}
\usepackage{bm}

\preprint{submitted to the Physical Review B}

\begin{document}

\title{Orbital magnetism in the half-metallic Heusler alloys}

\author{I. Galanakis}\email{I.Galanakis@fz-juelich.de}
\affiliation{Institut f\"ur Festk\"orperforschung,
Forschungszentrum J\"ulich, D-52425 J\"ulich, Germany}

\date{\today}

\begin{abstract}
Using the fully-relativistic screened Korringa-Kohn-Rostoker
method I study the orbital magnetism in the half-metallic Heusler alloys. 
Orbital moments are almost completely
quenched and they are negligible with respect to the spin moments.
The change in the atomic-resolved orbital moments can be easily
explained in terms of the spin-orbit strength and hybridization
effects. Finally I discuss the orbital and spin moments derived from
X-ray magnetic circular dichroism experiments.
\end{abstract}

\pacs{71.20.Be, 71.20.Lp, 75.50.Cc}
%71.20.-b   Electron density of states and band structure of crystalline solids
%71.20.Be   Transition metals and alloys
%71.20.Lp   Intermetallic compounds
%75.50.Cc   Other ferromagnetic metals and alloys
\maketitle

\noindent \textit{Introduction.} Half-metallic ferromagnets
consist a new class of materials which attracted a lot of
attention due to their possible applications in
spintronics.\cite{Zutic2004} In these materials the two spin bands
have a completely different behavior. While the majority spin band
(referred also as spin-up band) shows the typical metallic
behavior, the minority spin band (spin-down band) is
semiconducting. The spinpolarization at the Fermi level is 100\%
and these compounds could maximize the efficiency of the
magnetoelectronic devices.\cite{deBoeck2002}

de Groot and collaborators were the first to predict the existence
of half-metallicity. They
calculated the electronic structure of the Heusler alloy NiMnSb
and showed that the Fermi level in the minority band falls within
a gap while the majority band was metallic.\cite{deGroot1983}
Since then a lot of materials have been predicted to be
half-metals: other half-Heusler alloys (\textit{e.g}
PtMnSb),\cite{Galanakis2002a,Zhao} a large number of the full-Heusler
alloys (\textit{e.g} Co$_2$MnGe),\cite{Galanakis2002b,Ishida} the quaternary
Heusler alloys,\cite{Quaternary,QuaternaryB} some oxides 
(\textit{e.g} CrO$_2$ and Fe$_3$O$_4$),\cite{Soulen98} the manganites
(\textit{e.g} La$_{0.7}$Sr$_{0.3}$MnO$_3$),\cite{Soulen98}  
the double perovskites (\textit{e.g.} Sr$_2$FeReO$_6$),\cite{Kato}
the pyrites (\textit{e.g} CoS$_2$),\cite{Pyrites} the
transition metal chalcogenides (\textit{e.g} CrAs) and pnictides 
(\textit{e.g} CrSe) in the
zinc-blende or wurtzite structures,\cite{GalanakisZB,Xie,Akinaga}
the europium chalcogenides (\textit{e.g} EuS)\cite{Temmerman} and the diluted
magnetic semiconductors (\textit{e.g} Mn impurities in Si or
GaAs).\cite{FreemanMnSi,Akai98} Heusler alloys are particularly
interesting among these materials due to their very high Curie
temperature, which can attend 1000 K in the case of Co$_2$MnSn, and
the similarity between their crystal structure and the zinc-blende
structure adopted by the III-VI and IV-V binary semiconductors like
GaAs or ZnS.\cite{landolt}

Several papers have been devoted to the calculation of the
electronic structure of the half-metallic Heusler alloys. All these 
studies produced a similar description of their magnetic 
properties.\cite{calculations,Ishida,Picozzi} In 2002
Galanakis et al. have shown that the appearance of the gaps in
these alloys is directly connected to the magnetic spin moments
and moreover that the total spin magnetic moment $M_t$ scales
linearly with the total number of valence electrons $Z_t$
following the low $M_t=Z_t-18$ for the half-Heusler alloys like
NiMnSb and $M_t=Z_t-24$ for the full Heusler alloys like
Co$_2$MnGe.\cite{Galanakis2002a,Galanakis2002b} The orbital magnetic
moments of these alloys on the other hand have attracted much less
attention and results are scarce. Also experimentally only in few
cases the orbital magnetic moments have been determined via the
X-ray magnetic circular dichroic (XMCD) spectra of thin
films.\cite{Kimura1997,Felser} In this contribution I will present
a study of the atomic-resolved orbital magnetic moments of several
Heusler alloys using first-principles calculations.

\noindent\textit{Calculations Details.} To calculate the orbital
and spin magnetic moments I used the fully relativistic (FR) version of
the Korringa-Kohn-Rostoker (KKR) multiple-scattering Green
function method where the Dirac equation for the cell-centered
potentials in the atomic spheres (ASA) is solved.\cite{Pap02}
The Vosko, Wilk and Nusair parameterization\cite{Vosko} of the
local density approximation (LDA) is used for the exchange and
correlation potential. This method has been already employed to
calculate the effect of the spin-orbit coupling on the minority
band gap in the case of half-metallic
ferromagnets.\cite{Mavropoulos04} In the case of NiMnSb and
similar half-Heusler alloys it was shown that the spin-orbit
induces states within the gap but the effect is very weak and the
alloys show a region of very high spinpolarization
($\sim$99\%) instead of a gap; defects have a much more pronounced
effect on the destruction of the gap.\cite{defects} 

If I compare
the results obtained in this contribution by using the FR-KKR-ASA
with the results obtained in
Refs.~\onlinecite{Galanakis2002a} and \onlinecite{Galanakis2002b} using the
full-potential (FP) KKR method where the scalar-relativistic
approximation is employed (the spin-orbit coupling is not
taken into account), both versions of the KKR method reproduce a
similar description of the spin magnetic moments; the differences
are restricted to small deviations in the absolute values of the
spin magnetic moments. Both $C1_b$ and $L2_1$ structures of the
half- and full-Heusler alloys, respectively, are close-packed structures and ASA
is expected to give a good description of their electronic structure with respect to
FP. Moreover spin-orbit is a weak effect and only
marginally changes the spin moments.  
I should also note that LDA is known to underestimate  the orbital
moments by as much as 50\% but reproduces
the correct trends.\cite{orbital-magnetism,Grange}

\begin{table*}
\caption{\label{table1} Spin ($m_\mathrm{spin}$) and orbital
($m_\mathrm{orbit}$) magnetic moments in $\mu_B$ for the XMnSb
half-Heusler compounds. The last three columns are the total spin
and orbital magnetic moment and their sum, respectively}
\begin{ruledtabular}
\begin{tabular}{rrrrrrrrrr}
\multicolumn{10}{c}{MnSb-based half-Heusler alloys}\\ \hline &
$m^X_\mathrm{spin}$ & $m^X_\mathrm{orbit}$ &
$m^{Mn}_\mathrm{spin}$ & $m^{Mn}_\mathrm{orbit}$ &
$m^{Sb}_\mathrm{spin}$ & $m^{Sb}_\mathrm{orbit}$ &
$m^{total}_\mathrm{spin}$ & $m^{total}_\mathrm{orbit}$ &
$m^{total}$ \\ \hline FeMnSb  & -0.973&  -0.060&  2.943 &  0.034&
-0.040 &-0.002&  1.958  & -0.028&  1.930\\ CoMnSb  &-0.159&
-0.041& 3.201& 0.032& -0.101&  -0.001  &2.959  & -0.010  &2.949 \\
NiMnSb &0.245& 0.015 &3.720& 0.027  & -0.071&  -0.001 & 3.951  &
0.040 &3.991 \\ CuMnSb & 0.132  & 0.006 &  4.106   &0.032  & 0.028
& -0.006& 4.335  & 0.032   &4.367 \\ RhMnSb &-0.136 &-0.033 &3.627
&0.035& -0.141&  ~ -0& 3.360& 0.001& 3.361 \\ PdMnSb &0.067 &0.007
& 4.036& 0.028 & -0.117& ~ -0& 4.027 &0.035& 4.062 \\ AgMnSb
&0.106 &0.006& 4.334  & 0.031 &  0.040&   -0.007&  4.556 &  0.029
& 4.585 \\ IrMnSb &-0.201&  -0.094& 3.431& 0.092& -0.109 &-0.001
&3.130& -0.004 &3.126 \\ PtMnSb  &0.066& 0.006& 3.911& 0.057&
-0.086 &~ 0& 3.934& 0.063 &  3.997 \\ AuMnSb  &0.134 &  0.021 &
4.335 & 0.027 & 0.056& -0.006&  4.606&   0.044 &  4.650
\end{tabular}
\end{ruledtabular}
\end{table*}

\noindent \textit{Half-Heusler alloys containing Mn-Sb.} The first
family I will study is the MnSb-based half-Heusler alloys and in
Table \ref{table1} I have gathered their magnetic moments. To this
family belong the Fe-,Co-,Ni- and PtMnSb which are half-metallic
(HM). RhMnSb and IrMnSb are
isoelectronic to CoMnSb but the Fermi level falls within the
minority valence band and the HM is lost (the total spin moments are 
slightly above the ideal value of 3 $\mu_B$). 
The Cu-, Ag- and AuMnSb have 23 valence electrons
and if they were HM they should have a total spin moment of 5
$\mu_B$, but as it was shown in Ref.~\onlinecite{Galanakis2002a}
this value is practically impossible to get; it is energetically more
favorable to loose the HM. As a result also  the spin  moments of the Sb atoms 
are now parallel to the spin moments of the Mn atoms contrary to the other compounds.

The orbital moments are small with respect to the spin moments and
only in the case of IrMnSb the $m^{Ir}_\mathrm{orbit}$ approaches
the 0.1 $\mu_B$. In the case of the Sb atoms, the $sp$-bands lay low
in energy and are almost completely filled for both spin
directions.\cite{Galanakis2002a} There is a only a very small majority
spin $p$-weight around the Fermi level due to the antibonding $p$-$d$
hybrids. As a result the antimonium orbital moment is practically
zero for all compounds.

Mn atoms posses a large spin-magnetic moment in all Heusler alloys. The Mn spin-up
states are practically completely occupied while Mn admixture in
the occupied minority $d$ states is limited; it is mainly the X atom
which dominates the minority bonding $d$ states.\cite{Galanakis2002a} 
Mn orbital moment is less than 0.1$\mu_B$ is in all cases 
and remains parallel to the spin moment following the 3rd Hund rule. 
The latter rule, although derived for
atoms, stands also for solids with few exceptions.\cite{Openneer} It
states that if the $d$ band is more than half-filled (Mn has 7
$d$-electrons) then the spin and orbital moments should be
parallel. Increasing the valence of the X atom by one electron
either following the 3$d$ series
(Fe$\rightarrow$Co$\rightarrow$Ni$\rightarrow$Cu) or the 4$d$ series
(Rh$\rightarrow$Pd$\rightarrow$Ag) only scarcely changes the Mn orbital moment 
while there are significant variations  in the value of the Mn
spin moment.  If now  the X-atom changes along the 5$d$-elements series
(Ir$\rightarrow$Pt$\rightarrow$Au), the  increase of the Mn spin
moment by $\sim 0.5 \mu_B$ at every step is accompanied by a 
large decrease of the Mn orbital moment which is
practically halved. The increase of the spin moment is expected since 
the hybridization between Mn and a $d$ atom decreases as the valence of the $d$ atom
increases leading to a more atomiclike electronic structure around the Mn site.
The large effect on the Mn orbital moment in the case of the 5$d$ atoms has been 
already discussed in Ref.~\onlinecite{Openneer}, where using perturbation theory
it was shown that the large spin-orbit coupling of the
5$d$ elements has a large effect on the orbital moment of the 3$d$ neighboring
atoms in the case of alloys.

Finally for the X atom the orbital moment follows the Hund's
rules and is always parallel to the spin magnetic moment. Note that
the Fe, Co, Rh and Ir spin moments are antiparallel with respect
to the Mn atom. The orbital moment follows the changes of the spin
moment and it increases as the number of valence electrons
increase. As I  substitute Co for Fe the orbital moment increases
from -0.06 $\mu_B$ to -0.04 $\mu_B$ and then to 0.015 $\mu_B$  for
Ni in NiMnSb. The absolute value of the orbital moment depends strongly also
on the spin-orbit coupling. This is clearly seen if I  compare Ir
with Co. Both atoms have similar spin moments; -0.16 $\mu_B$  for
Co and -0.20 $\mu_B$  for Ir. On the other hand cobalt's orbital
moment is -0.04 $\mu_B$ while the Ir orbital moment is double as much
(-0.09$\mu_B$). Also hybridization plays an important role on the
value of the orbital moment, \textit{e.g} in FePt Fe has a spin
moment of 2.9 $\mu_B$ instead of -1.0 $\mu_B$ in FeMnSb but the
Fe orbital moment is similar in both cases; its absolute value is
0.07 $\mu_B$  for FePt and 0.06 $\mu_B$ for
FeMnSb.\cite{galanakisPMA}

Orbital moments from first-principle calculations exist for the
Ni-, Pd- and PtMnSb compounds obtained using the full-potential
linear muffin-tin orbitals method (FPLMTO).\cite{Galanakis2000}
While results for NiMnSb are similar to the present 
calculations this is not the case for the
Pd and Pt atoms in PdMnSb and PtMnSb compounds. FPLMTO predicts 
that their orbital moment is
antiparallel to the spin moment contrary to the present
calculations. This difference can arise from the treatment of the
spin-orbit coupling. Whilst in the present calculations the Dirac
equations are solved, in the case of the FPLMTO study  the spin-orbit
coupling is treated as a perturbation and since orbital moments
are very small this can lead to  such small deviations.

\begin{table*}
\caption{\label{table2}Spin ($m_\mathrm{spin}$) and orbital
($m_\mathrm{orbit}$) magnetic moments in $\mu_B$ for the X$_2$YZ
full-Heusler compounds. The last three colums are the total spin
and orbital magnetic moments and their sum, respectively}
\begin{ruledtabular}
\begin{tabular}{rrrrrrrrrr}
\multicolumn{10}{c}{Half-ferromagnetic full-Heusler alloys}\\
\hline & $m^X_\mathrm{spin}$ & $m^X_\mathrm{orbit}$ &
$m^Y_\mathrm{spin}$ & $m^Y_\mathrm{orbit}$ & $m^Z_\mathrm{spin}$ 
$m^Z_\mathrm{orbit}$ & $m^{total}_\mathrm{spin}$ &
$m^{total}_\mathrm{orbit}$ & $m^{total}$ \\ \hline Co$_2$MnAl&
0.745 &  0.012 &  2.599 &0.013& -0.091 & ~ -0 & 3.998& 0.038
&4.036
\\ Co$_2$MnSi & 0.994 &0.029& 3.022& 0.017 &  -0.078 & 0.001 &
4.932& 0.076 &  5.008 \\ Co$_2$MnGe&
0.950 &0.030 &3.095 &  0.020 &  -0.065 & 0.001 &  4.931 &0.081  & 5.012 \\
Co$_2$MnSn &0.905& 0.038 &3.257  & 0.025 &  -0.079 & ~  0 &4.988
&0.101 &5.089 \\ Co$_2$CrAl& 0.702& 0.012 & 1.644 &  0.008
&-0.082& ~ 0 &2.966 &  0.033& 2.999
\\  Co$_2$FeAl& 1.094& 0.045  & 2.753 &  0.060 &
-0.095& ~ -0 &4.847 &0.149 &4.996 \\  Fe$_2$MnAl &-0.311 & -0.015
& 2.633& 0.014 &-0.016& 0.001& 1.994 &-0.014 &1.980 \\ Mn$_2$VAl &
-1.398
&-0.034 &0.785& -0.009& 0.013& 0.005& -1.998& -0.073 &-2.071 \\
Rh$_2$MnAl& 0.304& -0.011& 3.431 &0.034 &-0.037 &-0.001 &4.002&
0.011& 4.013
\end{tabular}
\end{ruledtabular}
\end{table*}

Finally it was shown in Ref.~\onlinecite{Ebert90} that the orbital
moment is proportional to the difference between the number of
states of majority and minority spin at the Fermi level:
$m_{orbit}\propto n^\uparrow(E_\mathrm{F}) -
n^\downarrow(E_\mathrm{F})$. In the case of the half-metallic
systems $n^\downarrow(E_\mathrm{F})$=0 and thus the total orbital
moment should be parallel to the total spin moment. This is
not the case always as can be seen in Table~\ref{table1}. In Ref. \onlinecite{Ebert90}
it was assumed    
that the $t_{2g}$ and
$e_g$ states are degenerate and the local DOS of  all atoms is a Lorentzian; thus
the  applicability of this relation is restricted.

\noindent \textit{Half-metallic full-Heusler alloys.} In the
second part of my study I will concentrate on the half-metallic
full-Heusler alloys and in Table \ref{table2} I have gathered my
results. The orbital moments are quite small like the
half-Heuslers. In all cases with the exception of Rh atom in
Rh$_2$MnAl the Hunds rules are obeyed; note that for V in
Mn$_2$VAl the spin and orbital moments are antiparallel since V
$d$ valence shell is less than half-filled. The orbital moments of
the $sp$ atoms (Z sites) are almost zero for all cases as  in the half-Heuslers

The Co$_2$Mn-Z type compounds are the most interesting since they
present the highest Curie temperature among the known
half-metals.\cite{landolt} The comparison between the Al and Si compounds, 
which have one valence electrons difference, reveals large changes in
their magnetic properties. The Co spin moment increases by nearly 0.25 $\mu_B$ and
the Co orbital moment follows this change since it is more than
double for the Si compound. The increase in the Mn spin moments is
proportionally smaller and so do the orbital moments. Substituting 
now Ge or Sn for Si, which are
isovalent systems, has only a weak effect on the spin moments. Co
spin moment slightly decreases while the Mn spin moment slightly
increases. For both atoms the orbital moments show a small increase with
the atomic number.

The next step is to substitute Cr for Mn in Co$_2$MnAl. Co spin
moment is not affected by this substitution and so does its
orbital moment. Thus the Co orbital moment is mostly induced by
the spin-orbit coupling at the Co moment and is insensitive to
hybridization with the neighboring sites. Cr moments on the other
hand have to account for the missing electron and are considerably
smaller than the Mn ones. Substituting Fe for Mn in  Co$_2$MnAl has a 
more pronounced effect. Co spin moment increases by 0.35$\mu_B$ while its 
orbital moment is more than tripled. Its also interesting 
to compare  Co$_2$FeAl to the isoelectronic Co$_2$MnSi. Co spin moment in the case of
 Co$_2$FeAl is slightly larger   while the Co orbital moment is increased by
$\sim$50\%.

Comparing Co$_2$MnAl with Fe$_2$MnAl reveals only small
changes at the Mn site and the decrease in the total number for
valence electrons is taken care by Fe atoms. Substituting now Rh
for Co in the same compound leads to an increase of both the spin
and orbital moments of Mn since the hybridization between Mn and
Rh $d$ states is considerably smaller than between the Mn and Co
$d$ states. Finally I also calculated the properties of Mn$_2$VAl.
The increased hybridization between the Mn and its neighboring Mn
and V atoms leads to a large orbital moment at the Mn site although
its spin moment is halved with respect to the cases above where Mn
occupied the Y site.

To my knowledge calculations of the orbital moment exist only by
Picozzi et al.\cite{Picozzi} for the Co$_2$Mn-Si, -Ge and -Sn
compounds. The orbital moment at the Co site was found to be
around 0.02$\mu_B$ and at the Mn site around 0.008$\mu_B$. These
moments are slightly smaller than my values. The differences can
arise from the treatment of the spin-orbit coupling as
perturbation in their calculations.

\noindent \textit{Experiments.} Few experiments dedicated to the 
orbital magnetism exist on these
compounds. These experiments involve the obtaining of the XMCD spectra of
thin films. XMCD is the difference
between  the absorption spectra for left- and right-circular
polarized light involving  2$p$ core states excitations towards
unoccupied $d$ states. Elmers and collaborators\cite{Felser}  derived
orbital moments of 0.12 $\mu_B$ for Co, 0.04 $\mu_B$ for Cr and
0.33 $\mu_B$ for Fe in the case of a Co$_2$Cr$_{0.6}$Fe$_{0.4}$Al
thin film. If I compare these values with my calculations for the
Co$_2$CrAl and Co$_2$FeAl compounds they are one order of
magnitude larger. LDA usually gives orbital moments only halved
with respect to experiments.\cite{Grange} Also the XMCD derived spin
moments are half of the theoretical predicted values.
On the other hand Kimura et \textit{et al.}\cite{Kimura1997} studied the
NiMnSb and PtMnSb films and found that
$m_\mathrm{orbit}^{total}/m_\mathrm{spin}^{total}<0.05$ while in
my calculations this ratio is around 0.01. The spin moments derived by Kimura
\textit{et al.} experiments are also comparable
to the theoretical results. Thus the deviation   
between the present theoretical results and the experiments in Ref. \onlinecite{Kimura1997}
is considerably smaller than when compared to the ones in Ref.  \onlinecite{Felser}.

In both sets of experiments the orbital and spin moments are derived by
applying the sum rules to the XMCD spectra. The sum rules have  been derived using an
ionic model\cite{XMCD} and their application  to itinerant
systems, in particular to low symmetry systems, is strongly
debated\cite{chenwu} since XMCD probes mainly the region near
the surface of a film. Thus their application to experimental
spectra is not straightforward.  Elmer's and collaborators
sum-rule derived total spin moment is halved not only with respect to the 
theoretical results but most importantly also with respect to the value 
derived from the  SQUID measurements. This
inconsistency even between XMCD and SQUID measurements on the same sample 
shows that the application of sum rules to derive the
moments in the case of XMCD experiments on films is not really adequate.

 \noindent \textit{Summary.}  I have studied the orbital 
magnetism in the half-metallic Half-
and Full-Heusler alloys using the Dirac formalism within the
framework of the Korringa-Kohn-Rostoker Green's function method.
The quenching of the orbital moments is pretty complete and their
values are very small with respect to the spin moments. The change
in the atomic-resolved orbital moments can be easily explained in
terms of the spin-orbit strength and hybridization effects. 
Moments derived by applying the sum rules to the
experimental X-ray dichroic spectra of thin films should be treated 
with caution.

%%%%%%%%%%%%%%%%%%%%%%%%%%%%%%%%%%%%%%%%%%%%%%%%%%%%%%%%%%%%%%%%%


\begin{thebibliography}{99}

\bibitem{Zutic2004}
I. \v{Z}uti\'c, J. Fabian, and S. Das Sarma, Rev. Mod. Phys.
\textbf{76}, 323 (2004).

\bibitem{deBoeck2002}
J. de Boeck, W. van Roy, J. Das, V. Motsnyi, Z. Liu, L. Lagae, H.
Boeve, K. Dessein, and G. Borghs, Semicond. Sci. Tech.
\textbf{17}, 342 (2002).

\bibitem{deGroot1983}
R. A. de Groot. F. M. Mueller, P. G. van Engen, and K. H. J.
Buschow, Phys. Rev. Lett. \textbf{50}, 2024 (1983).

\bibitem{Galanakis2002a}
I. Galanakis, P. H. Dederichs, and N. Papanikolaou,  Phys. Rev. B
\textbf{66}, 134428 (2002).

\bibitem{Zhao}
M. Zhang \textit{et al.}, J. Phys: Condens. Matter \textbf{15}, 7891 (2003); M.
Zhang \textit{et al.}, J. Appl. Phys. \textbf{95}, 7219 (2003).

\bibitem{Galanakis2002b}
I. Galanakis, P. H. Dederichs, and N. Papanikolaou, Phys. Rev. B
\textbf{66}, 174429 (2002).

\bibitem{Ishida}
S. Fujii, S. Ishida, and S. Asano, J. Phys. Soc. Jpn. \textbf{64},
185 (1995); S. Ishida, S. Fujii, S. Kashiwagi, and S. Asano, J.
Phys. Soc. Jpn. \textbf{64}, 2152 (1995).

\bibitem{Quaternary}
Y. Miura, K. Nagao, and M. Shirai, Phys. Rev B \textbf{69}, 144413
(2004).

\bibitem{QuaternaryB}
I. Galanakis,  J. Phys: Condens. Matter \textbf{16}, 3089 (2004).

\bibitem{Soulen98}
R. J. Soulen Jr. \textit{et al.}, Science {\bf 282}, 85 (1998).

\bibitem{Kato}
H. Kato, T. Okuda, Y. Okimoto, Y. Tomioka, K. Oikawa, T. Kamiyama, and 
Y. Tokura, Phys. Rev. B \textbf{69}, 184412 (2004).

\bibitem{Pyrites}
T. Shishidou, A. J. Freeman, and R. Asahi,
Phys. Rev. B \textbf{64}, 180401 (2001).

\bibitem{GalanakisZB}
I. Galanakis, Phys. Rev. B \textbf{66}, 012406 (2002); I.
Galanakis and Ph. Mavropoulos, Phys. Rev. B \textbf{67}, 104417
(2003).

\bibitem{Xie}
S. Sanvito  and N. A. Hill, Phys. Rev. B \textbf{62}, 15553
(2000); B. Sanyal,  L. Bergqvist, and O. Eriksson, Phys. Rev. B
\textbf{68}, 054417 (2003);  W.-H. Xie, B.-G. Liu, and D. G.
Pettifor, Phys. Rev. B \textbf{68}, 134407 (2003); M. Zhang
\textit{et al.}, J. Phys: Condens. Matter \textbf{15}, 5017
(2003); J.-C. Zheng and J. W. Davenport, Phys. Rev. B \textbf{69},
144415 (2004).

\bibitem{Akinaga}
H. Akinaga, T. Manago, and M. Shirai, Jpn. J. Appl. Phys. 
\textbf{39}, L1118 (2000); M. Shirai,  J. Appl. Phys. \textbf{93}, 6844
(2003).

\bibitem{Temmerman}
M. Horne, P. Strange, W. M. Temmerman, Z. Szotek, A. Svane, and H. Winter,
J. Phys.: Condens. Matter \textbf{16}, 5061 (2004).

\bibitem{FreemanMnSi}
A. Stroppa, S. Picozzi, A. Continenza, and A. J. Freeman,
Phys. Rev. B \textbf{68}, 155203 (2003).

\bibitem{Akai98}
H. Akai, Phys. Rev. Lett. \textbf{81}, 3002 (1998).

\bibitem{landolt}
P. J. Webster and K. R. A. Ziebeck, in {\em Alloys and Compounds
of d-Elements with Main Group Elements. Part 2.}, edited by H. R.
J. Wijn, Landolt-Bo\"ornstein, New Series, Group III, Vol. 19/c
(Springer, Berlin), 1988, pp. 75-184.

\bibitem{calculations}
E. Kulatov and I. I. Mazin, J. Phys.: Condens. Matter {\bf 2}, 343
(1990); S. V. Halilov and E. T. Kulatov, J. Phys.: Condens. Matter
{\bf 3}, 6363 (1991);  S. J. Youn and B. I. Min,
Phys. Rev. B {\bf 51}, 10~436 (1995); V. N. Antonov, P. M.
Oppeneer, A. N. Yaresko, A. Ya. Perlov, and T. Kraft, Phys. Rev. B
{\bf 56}, 13~012 (1997).

\bibitem{Picozzi}
S. Picozzi, A. Continenza, and A. J. Freeman,  Phys. Rev. B
\textbf{66}, 094421 (2002).

\bibitem{Kimura1997}
A. Kimura, S. Suga, T. Shishidou, S. Imada, T. Muro, S. Y. Park,
T. Miyahara, T. Kaneko, and T. Kanomata, Phys. Rev. B \textbf{56},
6021 (1997).

\bibitem{Felser}
H. J. Elmers \textit{et al.}, Phys. Rev. B \textbf{67}, 104412
(2003).

\bibitem{Pap02}
N. Papanikolaou, R. Zeller, and P. H. Dederichs, J. Phys.: Condens.
Matter \textbf{14}, 2799 (2002).

\bibitem{Vosko}
S. H. Vosko, L. Wilk, and N. Nusair, Can. J. Phys. \textbf{58}, 1200
(1980).

\bibitem{Mavropoulos04}
Ph. Mavropoulos, K. Sato, R. Zeller, P. H. Dederichs, V. Popescu,
and H. Ebert, Phys. Rev. B \textbf{69}, 054424 (2004); Ph.
Mavropoulos,  I. Galanakis, V. Popescu, and P. H.Dederichs, J.
Phys.: Condens. Matter, in press.

\bibitem{defects}
D. Orgassa, H. Fujiwara, T. C. Schulthess, and W. H. Butler, Phys.
Rev. B \textbf{60}, 13237 (1999); S. Picozzi, A. Continenza, and
A. J. Freeman, Phys. Rev. B \textbf{69}, 094423 (2004).

\bibitem{orbital-magnetism}
H Ebert, Rep. Prog. Phys. \textbf{59},  1665 (1996).

\bibitem{Grange}
W. Grange, I. Galanakis, M. Alouani, M. Maret, J.-P. Kappler, and
A. Rogalev,  Phys. Rev. B \textbf{62}, 1157 (2000).

\bibitem{Openneer}
I. Galanakis, P. Ravindran, P.M. Oppeneer, L. Nordstr\"{o}m, P.
James, M. Alouani,  H. Dreyss\'e, and O. Eriksson, Phys. Rev. B
\textbf{63}, 172405 (2001).

\bibitem{galanakisPMA}
I. Galanakis,  M. Alouani and H. Dreyss\'e, Phys. Rev. B
\textbf{62}, 6475 (2000).

\bibitem{Galanakis2000}
I. Galanakis, S. Ostanin, M. Alouani, H. Dreyss\'e, and J. M.
Wills, Phys. Rev. B \textbf{61}, 4093 (2000).

\bibitem{Ebert90}
H. Ebert,     R. Zeller, B. Drittler, and P. H. Dederichs,
 J. Appl. Phys. \textbf{67}, 4576 (1990).

\bibitem{XMCD}
P. Carra, B. T. Thole, M. Altarelli, and X. Wang, Phys. Rev. Lett.
\textbf{70}, 694 (1993); B. T. Thole, P. Carra, F. Sette, and G.
van der Laan, Phys. Rev. Lett. \textbf{68}, 1943 (1992).

\bibitem{chenwu}
C. T. Chen \textit{et al.},  Phys. Rev. Lett.
\textbf{75}, 152 (1995); R. Wu, D. Wang, and A. J. Freeman, Phys.
Rev. lett. \textbf{71}, 3581 (1993); R. Wu and A.J. Freeman, 
\textit{ibid} \textbf{73}, 1994 (1994).

\end{thebibliography}
\end{document}